# The Important Electron Impact Collision Cross Sections with Methane

Abdelatif Gadoum[*], Djilali Benyoucef, Rachid Taleb

*Laboratoire Génie Electrique et Energies Renouvelables, Chlef University, Algeria*
[*]Email: gdabdelatif@gmail.com

*Abstract*—The electrical discharge plasma in methane is used in many fields of technology and the knowledge of the electron impact cross sections with the molecule of this gas is necessary for the modeling, which represents a good tool for understanding this phenomenon. The main object of this work is a finding the important electron impact cross sections with the methane molecule through the reviewing the theoretical and measured data found in the literature.

*Keywords-elastic collision; inelastic collision; electron impact cross section; methane molecule*

## I. INTRODUCTION

View of recent applications of the cold non-equilibrium thermodynamic plasma in different fields such as medicine and semiconductor manufacturing etc. [1-6]; researchers attention is attracted, experimental and theoretical studies are conducted as [7-20]. The study and the modeling of plasma require the knowledge of the physical processes; these last are based on the elementary microscopic mechanisms such as ionization, excitation, attachment, recombination, dissociation, and elastic collisions. Indeed, it is known that the knowledge of the cross sections of these mechanisms is indispensable to obtain the electron energy distribution function EEDF, the electron transport coefficients, and the rates of the different reactions. In this work we interest to accumulate the cross sections of electronic impact collision with methane found in the literature. Electron collisions with methane are especially important in plasmas. Methane plays a dominant role in the edge plasma region of high temperature plasma apparatus such as a tokamak. The methane $CH_4$ is the simplest poly-atomic hydrocarbon, and is present in the atmosphere of most planets and the interstellar surface [21-23]. Low-pressure radiofrequency plasma methane is used in many fields of technology [24-26], this molecular gas is considered to be a good test gas [27], several studies that were experimentally interested for example [28-33], or by numerical (analytical) study by mixing it with other gases such as $H_2$, $N_2$, etc. [34-37], or pure [38-42].

## II. ELECTRON CROSS SECTION FOR METHANE

When the molecule contains at least three atoms, the number of dissociative excitation channels, ionization and recombination increases drastically. For example, electron impact on the $H_2$ molecule yields only two fragments, and for $CH_4$ can produce $CH_3 + H$, $CH_2 + 2H$ (or $H_2$), $CH + 3H$ (or $H + H_2$), $C + 2H_2$ (or $2H + H_2$ ), where the atomic and the molecular products both being in excited states (electronic and/or vibratory). The processes of molecular fragmentation lead to a multiplication of molecular species in the plasma and to chains of long reactions before the complete dissociation of the initial molecule is accomplished. The methane molecule is relatively stable and it has four vibrational excitation modes as given in the table I.

TABLE I. Vibrational modes and excitation energies for $CH_4$ [43]

| | Mode | Energy |
|---|---|---|
| $\upsilon_1$ | Symmetric stretching | 0.362 |
| $\upsilon_2$ | Twisting | 0.190 |
| $\upsilon_3$ | Asymmetric stretching | 0.374 |
| $\upsilon_4$ | Scissoring | 0.162 |

Unfortunately to this day, it there are no effective electron dissociation cross sections of methane to the following channels; (1): $CH_3+H$, (2): $CH_2+H_2$, (3): $CH_2+2H$, (4): $CH+H_2+H$. The important electron-$CH_4$ molecule collisions are resumed in table 2.

TABLE II. Electronic impact reactions with threshold energy

| Process | Collision | Threshold (eV) |
|---|---|---|
| Elastic | $e+ CH_4 \Rightarrow e + CH_4$ | --- |
| Rotational | $e+ CH_4 \Rightarrow e + CH_4$ | 0.0078 |
| | $e+ CH_4 \Rightarrow e + CH_4$ | 0.013 |
| Vibrational | $e+ CH_4 \Rightarrow e + CH_4$ | 0.162 |
| | $e+ CH_4 \Rightarrow e + CH_4$ | 0.362 |
| Dissociation | $e+ CH_4 \Rightarrow e + CH_3 + H$ | 8.8 |
| | $e+ CH_4 \Rightarrow e + CH_2 + H_2$ | 9.4 |
| | $e+ CH_4 \Rightarrow e + CH + H_2 + H$ | 12.5 |
| | $e+ CH_4 \Rightarrow e + C + 2 H_2$ | 14 |
| Ionization | $e+ CH_4 \Rightarrow 2e + CH_4^+$ | 12.63 |
| Dissociative Ionization | $e+ CH_4 \Rightarrow 2e + CH_3^+ + H$ | 14.25 |
| | $e+ CH_4 \Rightarrow 2e + CH_2^+ + H_2$ | 15.1 |
| | $e+ CH_4 \Rightarrow 2e + CH^+ + H_2 + H$ | 19.9 |
| | $e+ CH_4 \Rightarrow 2e + C^+ + 2 H_2$ | 19.6 |
| | $e+ CH_4 \Rightarrow 2e + H_2^+ + CH_2$ | 20.1 |
| | $e+ CH_4 \Rightarrow 2e + H^+ + CH_3$ | 18.0 |

Mi-Song Yong et al. [43] confirm that for the dissociative there is no practical measure that distinguishes between (2) and (3) and no direct estimate of (4). Erwin and Kunc. [44], [45] presented a semi-empirical method to estimate the dissociative cross sections, but without including the dissociation to $C+2H_2$ (or $2H+H_2$). Figure 1 represents the cross sections of electron impact collision with methane for: momentum transfer cross section, rotational excitation cross section for the energies (0.0078 eV and 0.013 eV) eV, vibrational excitation cross section for the energies (0.162 eV and 0.362 eV), the dissociation excitation for the channels: $CH_3+H$, $CH_2+H_2$, $CH+H_2+H$, $C+2H_2$, the ionization into $CH_4^+$, and the dissociative ionization cross section for the channels: $H+CH_3^+$, $CH_2^+ + H_2$, $CH^+ + H_2 + H$, $C^+ + 2H_2$, $H_2^+ + CH_2$, $H^+ + CH_3$.

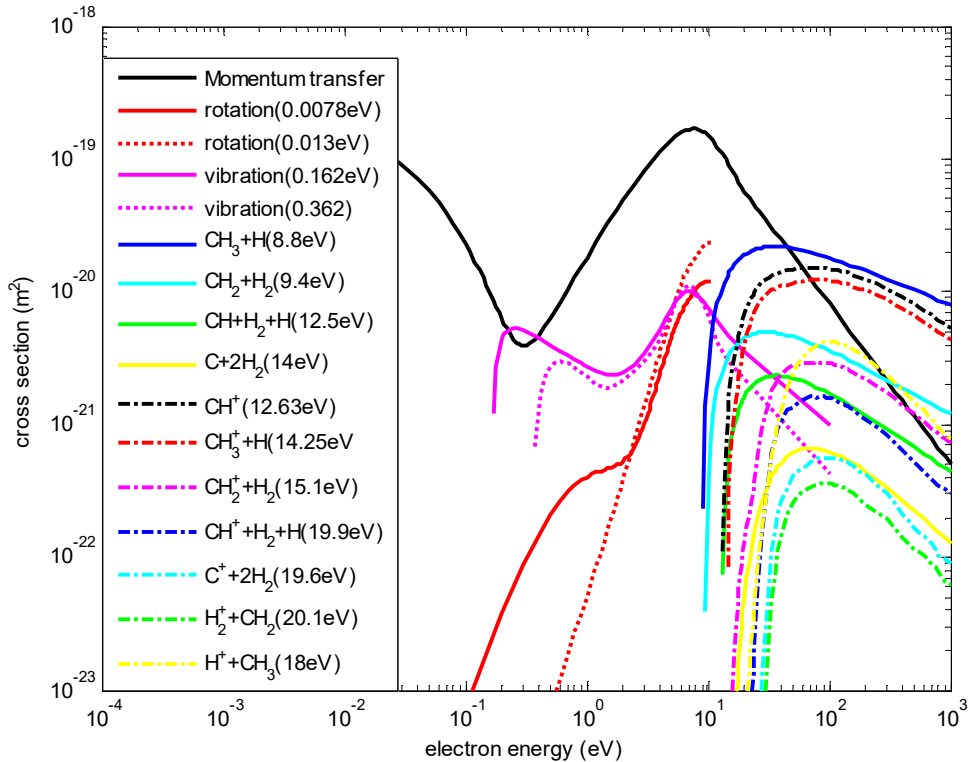

Fig.1: electron impact collision cross sections with methane [25]

### III. CONCLUSION

In this study a complete set of electronic impact collision cross sections with methane have been reported. These cross sections are very important for the study of the electrical discharge plasma in methane gas, and can be used especially in the modeling as input data for particle models.